# Lithium-salt-based deep eutectic solvents: Importance of glass formation and rotation-translation coupling for the ionic charge transport

A. Schulz, P. Lunkenheimer[a], and A. Loidl

**AFFILIATIONS**

Experimental Physics V, Center for Electronic Correlations and Magnetism, University of Augsburg, 86135 Augsburg, Germany

[a]Author to whom correspondence should be addressed: peter.lunkenheimer@physik.uni-augsburg.de

**ABSTRACT**

Lithium-salt-based deep eutectic solvents, where the only cation is Li$^+$, are promising candidates as electrolytes in electrochemical energy-storage devices like batteries. We have performed broadband dielectric spectroscopy on three such systems, covering a broad temperature and dynamic range that extends from the low-viscosity liquid around room temperature down to the glassy state approaching the glass-transition temperature. We detect a relaxational process that can be ascribed to dipolar reorientational dynamics and exhibits the clear signatures of glassy freezing. We find that the temperature dependence of the ionic dc conductivity and its room-temperature value also are governed by the glassy dynamics of these systems, depending, e.g., on the glass-transition temperature and fragility. Compared to the previously investigated corresponding systems, containing choline chloride instead of a lithium salt, both the reorientational and ionic dynamics are significantly reduced due to variations of the glass-transition temperature and the higher ionic potential of the lithium ions. These lithium-based deep eutectic solvents partly exhibit significant decoupling of the dipolar reorientational and the ionic translational dynamics and approximately follow a fractional Debye-Stokes-Einstein relation, leading to an enhancement of the dc conductivity, especially at low temperatures. The presented results clearly reveal the importance of decoupling effects and of the typical glass-forming properties of these systems for the technically relevant room-temperature conductivity.

## I. INTRODUCTION

Deep eutectic solvents (DES) have come into the focus of recent research as their properties make them promising candidates for numerous applications, e.g., in material synthesis or electrochemical devices.[1,2,3,4,5,6,7,8,9,10,11] Most DESs are easy to produce, sustainable, and biocompatible, partly being even composed of constituents found in nature. Thus, they turn out to be "greener" alternatives to ionic liquids, considered for similar applications.

As in all eutectics, in DESs a melting-point reduction arises due to the mixing of two or more components, making them liquid at room temperature. In the most common class of DESs, these components are molecular hydrogen-bond donors (HBD) like glycerol or urea, mixed with a salt acting as hydrogen-bond acceptor. The latter commonly is a quaternary ammonium salt, like in the often-investigated DESs glyceline, ethaline, and reline, which are mixtures of choline chloride with glycerol, ethylene glycol, or urea, respectively, all with 1:2 molar ratio. Their large salt fraction leads to considerable ionic conductivity, a prerequisite for electrochemical applications. However, for these applications, the most prominent one being electrolytes in batteries, usually the presence of ions like Li$^+$ or Na$^+$ is required. In DESs as those discussed above, this can be achieved by admixing salts like lithium bis(trifluoromethane)sulfonimide (LiTFSI) or LiPF$_6$.[9]

An alternative approach is the investigation of DESs where the only cation is Li$^+$ or Na$^+$. This avoids the possible accumulation of the large cations (e.g., choline$^+$) at the electrode, blocking it for the Li$^+$ ions.[12] For example, in a molecular-dynamics study, urea mixed with LiTFSI was proposed to be a promising DES for electrochemical applications.[13] Indeed, in an earlier work,[14] relatively high dc conductivities were detected in this system. The existence of Li-salt-based DESs was also reported for the common HBDs glycerol and ethylene glycol, e.g., in Refs. 15 and 16.

The application of DESs as electrolytes in electrochemical devices requires a high ionic dc conductivity, $\sigma_{dc}$, beyond about $10^{-4}$ $\Omega^{-1}$cm$^{-1}$ at room temperature. To optimize the conductivity and to understand the considerable variation of $\sigma_{dc}$ of different DESs,[5,9,12,15,23] a better knowledge of the ionic and molecular motions within these materials is desirable. For this purpose, dielectric spectroscopy is a well-suited experimental method as it can simultaneously provide information on the translational ionic and the reorientational molecular motions.[17,18] In DESs, the latter arise because the HBDs usually are asymmetric molecules with rotational degrees of freedom and in many cases, this is also the case for at least one of the added ion species. Several previous works have revealed that the reorientational molecular motions, present in certain ionic conductors, including ionic liquids and superionic and plastic crystals, seem to be highly relevant for ionic mobility.[19,20,21,22] Interestingly, consistent with this notion, for the DESs glyceline and ethaline, we recently found a close correlation of their ionic dc conductivity with the reorientational dynamics as detected by dielectric spectroscopy.[23] However, for reline some minor but significant deviations showed up.

Moreover, dielectric spectroscopy is also able to obtain valuable information on the glass transition, often occurring for eutectic mixtures at low temperatures. Indeed, many DESs seem to be glass-forming liquids[4,24] but, overall, this phenomenon is only rarely investigated in this material class. In a previous work,[23] we found that the temperature dependence of both the rotational and ionic dynamics of glyceline, ethaline, and reline indeed reveal the characteristics of glassy freezing. This also affects their room-temperature properties, e.g., the dc conductivity strongly depends on the glass-transition temperature.

In the present work, we report the results of broadband dielectric spectroscopy on three Li-salt-based DESs, with the same HBDs as the previously investigated[23,25] glyceline, ethaline, and reline. We especially treat the question of the possible coupling of the translational and reorientational dynamics in these systems and their glassy freezing. It should be noted that so far there are only few earlier dielectric investigations of DESs[16,23,24,26,27,28,29,30] often not addressing the reorientational dynamics, and, to our knowledge, none of them has tackled the technically relevant, purely Li-salt-based DESs.

## II. Experimental Details

The two salts LiTFSI and lithium triflate (LiOTf), as well as glycerol, were purchased from Sigma Aldrich, while urea and ethylene glycol are obtained from Alfa Aesar. The first DES, denoted LiTFSI/urea (LiTFSI + urea, 1:3.1 molar ratio), was prepared along the lines of Ref. 14 by blending appropriate amounts of the components inside a glass vessel and vacuum drying the mixture for 6.5 hours at 375 K. This way, any significant water entrapments were removed, while the source materials were thoroughly mixed. Similarly, LiOTf/Gly (LiOTf + glycerol, 1:3 molar ratio) and LiOTf/EG (LiOTf + ethylene glycol, 1:4 molar ratio) were produced, as described in Ref. 15, by heating the mixtures of the starting materials at 355 K for around 20 hours. All three DESs appeared as colorless, translucent, and homogeneous liquids with LiOTf/EG exhibiting a noticeably lower viscosity, compared to the other two solvents. The water content of all samples was tested by coulometric Karl-Fischer-Titration and determined to be 0.52, 0.006, and 0.21 wt% for LiOTf/Gly, LiOTf/EG, and LiTFSI/urea, corresponding to 3.0, 0.038, and 1.2 mol%, respectively. For LiOTf/Gly and LiOTf/EG, the water content is smaller than in Ref. 15 where 3.5 and 5 wt% were specified, respectively (no information on the water content in LiTFSI/urea is provided in Ref. 14). We want to point out that there are several works, explicitly reporting the influence of the water content on the electrical properties and viscosity of DESs.[30,31,32] They allow to conclude that for concentrations as in the present samples, the modifications of the relaxation properties and conductivity, compared to water-free samples, are minor and can be neglected. To prevent water uptake, during the measurements the samples were kept in a dry nitrogen atmosphere. To cover a broad frequency range from about 1 Hz to 1 GHz, the dielectric measurements were performed with two different techniques, frequency-response analysis ($\nu <$ 3 MHz) and coaxial reflectometry[33] ($\nu >$ 1 MHz). For more information, the reader is referred to Ref. 23. Furthermore, differential scanning calorimetry (DSC) measurements were conducted using a DSC 8500 from PerkinElmer at a scanning rate of 10 K/min during heating and cooling. The onset of the step-like increase of the heating data revealed the glass-transition temperatures of the three eutectic mixtures.

## III. RESULTS AND DISCUSSION

### A. Dielectric spectra

In the following, we adhere to the approach in our earlier works on DESs[23,25,34] and ionic liquids,[22] analyzing the complex dielectric permittivity, $\varepsilon^* = \varepsilon' - i\varepsilon''$, which is the natural representation for materials with reorientational degrees of freedom and also allows for a direct comparison with our previous results on glyceline, ethaline, and reline.[23,34] In addition, we provide the real part of the conductivity $\sigma'$, which is proportional to $\varepsilon''\nu$ and enables the direct identification of the dc conductivity. At the end of this section, we will also discuss an evaluation of the present experimental results in terms of the dielectric modulus.

Figure 1 shows spectra of the dielectric constant ($\varepsilon'$), loss ($\varepsilon''$) and conductivity ($\sigma'$) as obtained for LiOTf/Gly at various temperatures. Overall, these spectra are qualitatively similar to those reported for the corresponding choline-chloride-based DESs with the same HBDs.[23,34] The increase of $\varepsilon'(\nu)$ [Fig. 1(a)] to unreasonably high values, exceeding $10^6$, observed at low frequencies and high temperatures, is due to electrode polarization.[35] This non-intrinsic phenomenon is a common finding for ionic conductors. It also explains the low-frequency decrease of $\sigma'(\nu)$ detected at the highest temperatures [Fig. 1(c)],[36] and (due to the relation $\varepsilon'' \propto \sigma'/\nu$) the corresponding more shallow $\varepsilon''(\nu)$ trace, e.g., observed below about 100 Hz for the 286 K curve [Fig. 1(b)]. At frequencies beyond the electrode-dominated regime, the $\varepsilon'$ spectra reveal the typical indications of a dipolar relaxation process, signified by a step-like decrease with increasing frequency (see inset of Fig. 1 for a zoomed view). The points of inflection of these steps strongly shift to lower frequencies upon cooling, typical for the slowing down of the dynamics found in glass-forming materials.[18,37] At room temperature, the relaxation step in $\varepsilon'(\nu)$ is located at roughly 100 MHz. This is of similar order as reported for pure glycerol,[37] in agreement with the findings for glyceline.[23,38] Thus, reorientations of the glycerol molecules, representing 75% of the sample constituents, can be assumed to play a major role in the generation of the relaxation process observed in this DES. The second dipolar entity in this DES, LiOTf, does not lead to a separate relaxation process in the spectra and probably its rotations are closely coupled to those of glycerol. The detection of only a single relaxation process for mixtures of two types of dipoles is an often-found phenomenon.[39,40]

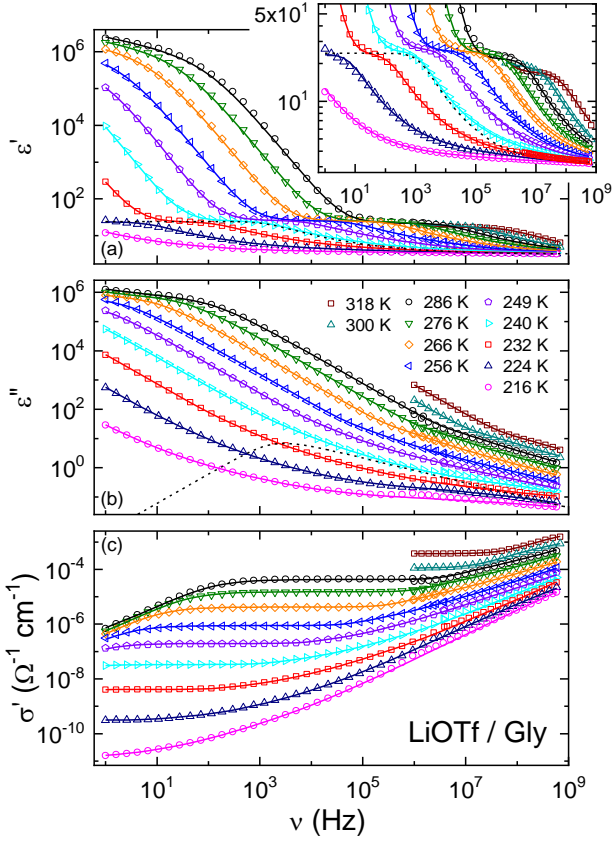

usually termed $\beta$ relaxations, which are common phenomena of dipolar glass-forming liquids,[42,43,44] including ionically conducting systems.[22,23,45] The discussion of these effects is out of the scope of the present work and these low-temperature spectra provide neither any information on the main reorientational process (termed $\alpha$ relaxation) nor on the dc conductivity.

The ionic dc conductivity of LiOTf/Gly is signified by the frequency-independent plateaus in $\sigma'(\nu)$ at frequencies beyond the mentioned electrode-dominated regime [Fig. 1(c)]. Its absolute value strongly decreases with decreasing temperature, mirroring the essentially thermally-activated nature of the ionic charge transport. At higher frequencies, the dc plateau crosses over into a region with increasing $\sigma'(\nu)$. This mirrors the mentioned relaxational response (the right flanks of the $\alpha$-relaxation peaks) as detected in $\varepsilon''(\nu)$, contributing to the conductivity spectra due to the close relation of both quantities ($\sigma' \propto \varepsilon'' \nu$).

**FIG. 1.** Frequency dependence of the dielectric constant $\varepsilon'$ (a), the dielectric loss $\varepsilon''$ (b), and the real part of the conductivity $\sigma'$ (c), measured at various temperatures for LiOTf/Gly. For the two highest temperatures, only measurements with the high-frequency device are provided ($\nu \geq 1$ MHz), which is sufficient to reveal the dc conductivity and dipolar relaxation. The inset shows a zoomed view of the intrinsic relaxation steps in $\varepsilon'(\nu)$. The solid lines in (a) and (b) are fits with an equivalent circuit, assuming a distributed RC circuit to describe the blocking electrodes,[35] intrinsic relaxation processes ($\alpha$ and secondary), and a contribution from dc-conductivity as described in the text. The real and imaginary parts of the permittivity were fitted simultaneously and the fits of the conductivity were calculated using $\sigma' = \varepsilon'' \varepsilon_0 \omega$. As an example, the dashed lines in (a), (b), and in the inset show the contribution of the $\alpha$ relaxation for 240 K.

In general, relaxation steps in $\varepsilon'(\nu)$ should be accompanied by relaxation peaks in $\varepsilon''(\nu)$. However, as often found for systems with relatively high conductivity, in the present case these peaks are partly superimposed by the strong dc-conductivity contribution in the loss spectra, $\varepsilon''_{dc} \propto \sigma_{dc}/\nu$, which leads to a $1/\nu$ increase towards lower frequencies. Therefore, only the right flanks of the expected peaks show up, which, at high frequencies, give rise to a shallower decrease of $\varepsilon''(\nu)$, compared to the dc contribution [cf. the dashed line in Fig. 1(b) indicating the unobscured relaxation peak as obtained from the fits described below]. A derivative analysis of $\varepsilon'(\nu)$ as proposed in Ref. 41 and previously performed for glyceline[34] supports this notion.

At the lowest temperature in Fig. 1, where the relaxation step in $\varepsilon'(\nu)$ has essentially shifted out of the frequency window, the loss spectrum in Fig. 1(b) reveals a region with weak frequency dependence at high frequencies (e.g., at $\nu > 10^5$ Hz for 216 K). This points to further minor contributions in this region, e.g., due to secondary processes,

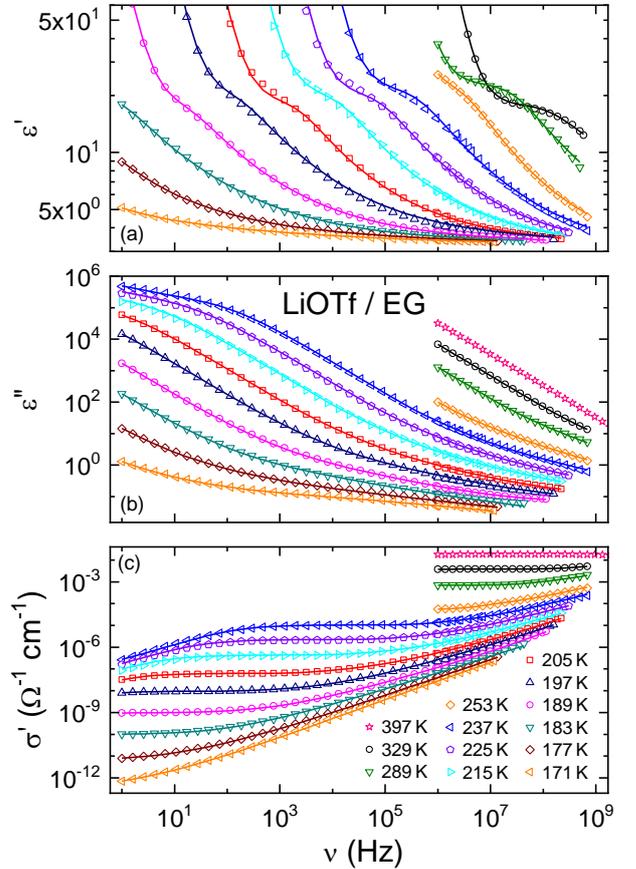

**FIG. 2.** Frequency dependence of the dielectric constant $\varepsilon'$ (a), the dielectric loss $\varepsilon''$ (b), and the real part of the conductivity $\sigma'$ (c) of LiOTf/EG [corresponding plot as in Fig. 1 but with frame (a) providing a zoomed view of the intrinsic $\alpha$ relaxation]. The solid lines have the same meaning as in Fig. 1. For 397 K, no meaningful results could be obtained for $\varepsilon'$, due to the high conductivity at this temperature.

Figures 2 and 3 show the dielectric spectra of LiOTf/EG and LiTFSI/urea.[46] The upper frames provide a zoomed view of the intrinsic relaxation steps in $\varepsilon'(\nu)$. With the typical signatures of electrode polarization, an intrinsic relaxation process in $\varepsilon'(\nu)$ and the dc plateau in $\sigma'(\nu)$, they qualitatively resemble those of LiOTf/Gly in Fig. 1. Especially for LiOTf/EG, at the lowest temperatures a weak broad loss peak or shoulder [e.g., around $10^4$ Hz in the 171 K curve in Fig. 2(b)] is unequivocally detected. This confirms the $\beta$-relaxation scenario discussed above to explain the nearly frequency-independent loss observed at low temperatures.

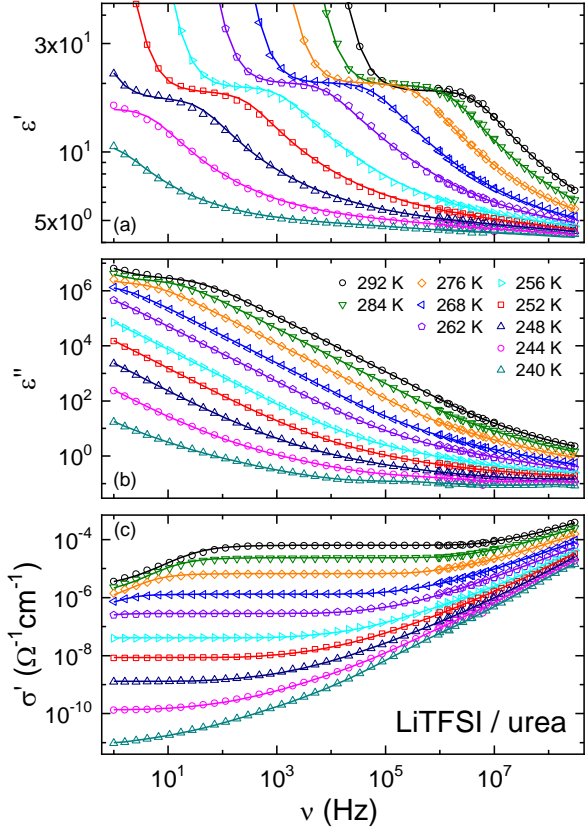

**FIG. 3.** Frequency dependence of the dielectric constant $\varepsilon'$ (a), the dielectric loss $\varepsilon''$ (b), and the real part of the conductivity $\sigma'$ (c) of LiTFSI/urea (corresponding plot as in Fig. 2). The solid lines have the same meaning as in Figs. 1 and 2.

To fit the data of Figs. 1-3, we formally describe the electrode polarization effects by a distributed RC circuit connected in series to the bulk sample,[35] as previously employed for DESs and other ionic conductors.[21,22,23,34] For the $\alpha$ relaxation, we use the Cole-Davidson (CD) formula[47] and the secondary relaxation is fitted by the Cole-Cole (CC) function,[48] both commonly-used empirical fit functions of $\alpha$ and secondary processes, respectively.[17,18,44] Finally, the dc-conductivity contribution to the loss is accounted for by $\varepsilon''_{dc} = \sigma_{dc}/(\varepsilon_0\omega)$ (with $\varepsilon_0$ the permittivity of free space and $\omega = 2\pi\nu$). The solid lines in Figs. 1-3 show fits with this approach, simultaneously performed for $\varepsilon'(\nu)$ and $\varepsilon''(\nu)$. Just as for the choline-chloride-based DESs,[23,34] the experimental spectra can be well fitted in this way. Only for LiTFSI/urea, we found that two CC functions had to be employed to properly fit the high-frequency part of the low-temperature curves at $T \leq 244$ K. We want to point out that, depending on temperature, only part of the different contributions had to be employed for the fits, thus avoiding an excessive number of fit parameters. For example, at low temperatures, the electrode effects can be neglected and at high temperatures, the secondary relaxations play no role.

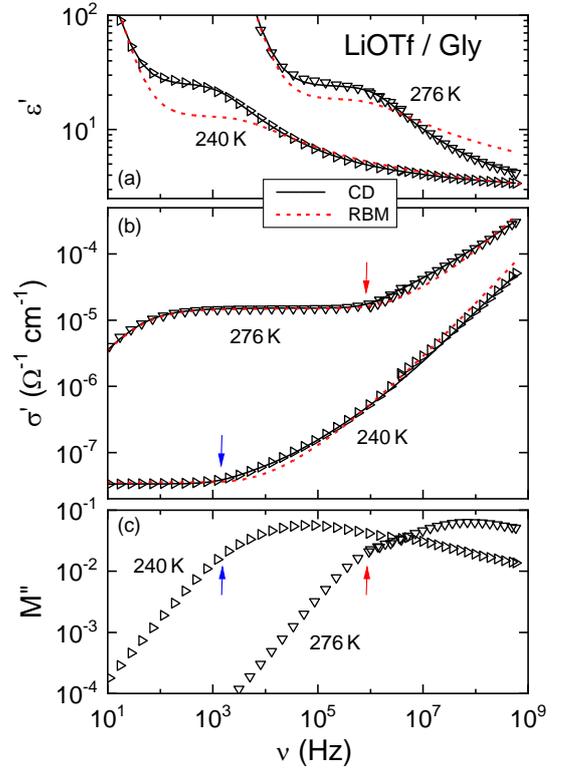

**FIG. 4.** Frequency dependence of $\varepsilon'$ (a), $\sigma'$ (b), and $M''$ (c) of LiOTf/Gly at two temperatures. The solid lines in (a) and (b) represent fits using a CD function for the $\alpha$-relaxation as already shown in Figs. 1(a) and (c). The dashed lines are fits using the RBM.[49] In both cases, the electrode polarization was modelled by a distributed RC circuit, connected in series to the bulk,[35] and for the lower temperature an additional secondary relaxation was included in the fits, modelled by the CC function. The dielectric constant and conductivity were fitted simultaneously. Please note the indication of a double-peak structure in the modulus representation (c), most obvious at 276 K. The arrows in (b) and (c) demonstrate that, instead of the main peaks, the low-frequency shoulders observed in the modulus spectra arise close to the crossover frequency in $\sigma'(\nu)$..

It seems worth mentioning that the above analysis departs from the canonical interpretation given in the majority of literature on the frequency-dependent conductivity in ionic conductors, where the mentioned crossover from the dc plateau in $\sigma'(\nu)$ into a region with increasing $\sigma'$ is interpreted as transition from the dc to ac conductivity. According to this interpretation, data as shown in Figs. 1-3 are analyzed in terms of purely translational charge transport via ion hopping, without assuming any contributions from dipolar reorientation dynamics. A prominent example is the random free-energy barrier hopping model (RBM),[49,50] which predicts a relaxation step in $\varepsilon'(\nu)$ that arises from local ion motions. It was previously applied, e.g., to two DES systems[24] and several ionic liquids.[51,52,53,54] Within this framework, the mentioned increase of $\sigma'(\nu)$ at high frequencies is due to ac conductivity arising from hopping charge transport. As an example for the application of this approach to the present data, Fig. 4 shows fits with the RBM prediction of the LiOTf/Gly spectra at two temperatures (dashed lines). In both cases, the same additional contributions as in the original CD fits shown in Fig. 1 were used, namely a distributed RC circuit to formally account for the electrode polarization and, for the lower temperature, a secondary relaxation process. For both temperatures, the quality of the RBM fits is significantly lower than for the fits assuming a reorientational relaxation process modeled by a CD function (solid lines). As argued in Refs. 23 and 34 and discussed above, it is clear that the well-known reorientational motions of the HBDs, making up 75 - 80 mol% of the investigated DESs, should lead to significant relaxational signatures in the spectra. As the RBM exclusively considers translational ion motions and was not conceived to account for dipolar reorientations, it cannot be expected to fit data governed by such dipolar relaxations. This is becoming especially obvious in typical DESs with their high content of strongly dipolar components. In other ionic conductors, like polymers or ionic liquids, where the dipolar relaxations are less clearly pronounced in relation to the dc conductivity, the RBM may provide reasonable fits but care has to be taken to clarify contributions from possible dipolar reorientation processes. Only for ionic conductors lacking any dipolar degrees of freedom like the ionic melt $[Ca(NO_3)_2]_{0.4}[KNO_3]_{0.6}$ (Ref. 55) its application is straightforward.

Concerning the present data, of course it cannot be excluded that contributions from ion hopping as described by the RBM, and dipolar reorientations superimpose in the measured data. However, as evidenced in our above discussion of Figs. 1-3 and as also found in our previous works,[23,34] the DES spectra can be explained without invoking any additional ac-conductivity contributions or relaxational response arising from ionic charge transport, whose only contribution to the spectra seems to be the dc conductivity. Therefore, in light of Occam's razor, here we describe our data exclusively by reorientational dynamics (which inevitably has to exist in these materials) and dc charge transport. In Ref. 34, the feasibility of an approach combining the RBM prediction and dipolar reorientation dynamics was demonstrated, but the increased number of parameters and problematic deconvolution of the different spectral contributions makes it difficult to arrive at definite conclusions in this way for the present samples.

Another way to present and analyze the result of dielectric measurements on ionically conducting systems is provided by the dielectric-modulus representation with the complex modulus $M^*$ being the inverse of the complex permittivity, i.e., $M^* = 1/\varepsilon^*$.[56] Although the applicability and interpretation of this representation is controversially discussed,[57,58,59,60] it is often used to analyze such data. One should be aware, however, that for systems with simultaneous reorientational and translational motions some problems may occur for this representation as discussed, e.g., in Ref. 34. In fact, in the pioneering work on the modulus formalism by Macedo *et al.*[56] it was noted that this approach "considers electrical dispersion phenomena in ionic conductors which contain no permanent molecular dipoles". In Fig. 5, as an example for the present data, we show spectra of the imaginary part of the dielectric modulus $M''$ for LiOTf/Gly, revealing peaks that shift to higher frequencies with increasing temperature, typical for relaxation processes. Within the modulus formalism,[56] peaks in $M''(\nu)$ are ascribed to a so-called conductivity relaxation and fitted using common empirical relaxation functions like the Cole-Davidson, Havriliak-Negami (HN), or Kohlrausch-Williams-Watts function.[56,61,62] The obtained relaxation times are then assumed to provide a measure for the ionic mobility. We found that the present $M''$ spectra at the lowest temperatures can be reasonably well fitted using the HN function (lines in Fig. 5). However, a closer look at Fig. 5 reveals significant, successively increasing deviations of fits and experimental data, occurring for the higher temperatures. This is due to a clear shoulder arising at the low-frequency flanks of the $M''$ peaks. Similar behavior was also found for the two other investigated DESs.

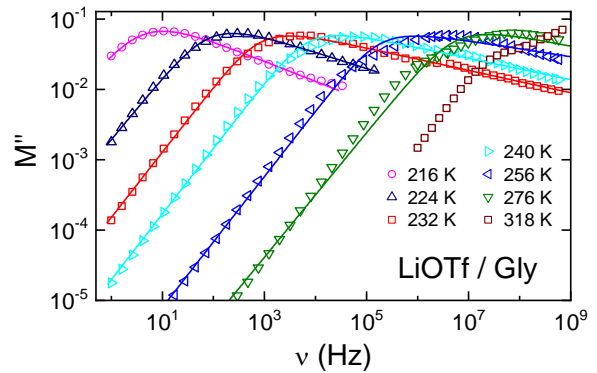

**FIG. 5.** Frequency dependence of the imaginary part of the dielectric modulus $M''$, measured at various temperatures for LiOTf/Gly. The lines show fits with the HN equation.

As discussed, e.g., in Refs. 34, 63, and 64, dipolar relaxation processes, which must be present in our DESs due to the high content of dipolar HBDs, not only produce peaks in the dielectric loss spectra but also in the dielectric modulus, where they show up at higher frequencies than in $\varepsilon''(\nu)$. Therefore, in such ionically conducting dipolar systems, two peaks arising from dipolar relaxation (at higher frequencies) and from the conductivity relaxation (at lower frequencies) should show up, as already treated in detail long ago by Johari and Pathmanathan.[63] This explains the low-frequency shoulders in the present spectra. As revealed by a direct comparison of $M''$ and $\sigma'$ in Figs. 4(b) and (c), these shoulders occur close to the mentioned crossover-frequency in $\sigma'(\nu)$ (cf. arrows in the figures), typical for conductivity relaxations. It is clear that spectra as in Fig. 5 do not allow for an unequivocal deconvolution of the conductivity from the dipolar relaxation and that the only directly observable peak frequency in $M''$ is not related to the ionic mobility. Moreover, in ionic conductors without significant reorientational relaxations, the $\sigma'(\nu)$ crossover is due to a transition from dc to ac conductivity. Then the crossover frequency, e.g., in the RBM, is believed to represent a characteristic time scale of ionic motion, also mirrored by the modulus peak frequency. In the present case, however, this crossover is due to a transition from dc conductivity to dipolar relaxation and the significance of the shoulder in $M''$ for the characterization of ionic mobility is questionable. For these reasons, we refrain here from an evaluation of the conductivity relaxation in our samples.

However, a brief discussion of the relaxation times determined from the main peaks in $M''(\nu)$ will be given below.

## B. Dc conductivity and $\alpha$-relaxation time

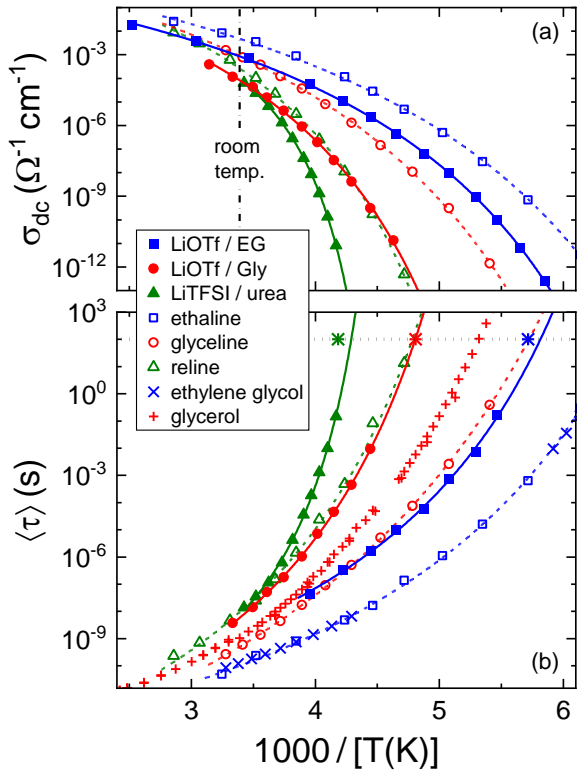

**FIG. 6.** Arrhenius plot of the dc conductivity (a) and mean $\alpha$-relaxation time (b) for the DESs investigated in the present work and in Ref. 23. In addition, in (b) also $\tau(T)$ data for the pure HBDs glycerol[82] (plusses) and ethylene glycol[23] (crosses) are included, as well as glass-transition temperatures for the lithium-salt-based DESs (stars) obtained via DSC measurements. The solid and dashed lines in (a) and (b) are fits with the VFT law, Eqs. (1) or (2), respectively. The vertical dash-dotted line in (a) indicates room temperature (295 K). The horizontal dotted line in (b) denotes $\tau(T_g) = 100$ s, which enables a rough estimate of the glass-transition temperature.

From an application point of view, the dc conductivity of these systems, reaching technically relevant values of $> 10^{-4}$ $\Omega^{-1}$cm$^{-1}$ at room temperature,[14,15,16] is the most interesting quantity. The temperature-dependent dc conductivities of the investigated DESs as resulting from the fits in Figs. 1-3 are shown in Fig. 6(a) using an Arrhenius representation (closed symbols). In all samples, we find clear deviations from a simple thermally-activated temperature dependence. This is typical for glass-forming ionic conductors[21,22,65] and mirrors the general non-Arrhenius behavior, characteristic of the structural dynamics of glass-forming liquids. While the $\sigma_{dc}(T)$ traces of the three investigated DESs approach each other at high temperatures, at low temperatures their dc-conductivity values deviate by many decades. As previously reported for ethaline, glyceline, and reline [open symbols shown for comparison in Fig. 6(a)[23]], for the HBD ethylene glycol the highest and for the

HBD urea the lowest conductivity is detected. These low-temperature deviations seem to be directly related to the different glass-transition temperatures $T_g$ of these systems determined by DSC, which are 175, 208, and 239 K for LiOTf/EG, LiOTf/Gly, and LiTFSI/urea, and 155, 175, and 205 K for ethaline, glyceline, and reline,[23] respectively. Having in mind the simple picture of particles diffusing through a viscous medium, it is reasonable that the strongly increasing viscosity $\eta(T)$ when approaching $T_g$,[66] accompanied by a slowing down of molecular dynamics, causes a reduction of the ionic mobility.

Another notable effect revealed by Fig. 6(a) are the systematic differences between the DESs containing lithium salts and those with the same HBDs but containing the choline-chloride salt instead (cf. closed and open symbols with the same color). In all three cases, the lithium systems exhibit significantly lower conductivities. These differences are most pronounced at low temperatures but, at least for LiOTf/EG and LiOTf/Gly, persist up to the highest investigated temperatures. As the Li$^+$ ions are by far smaller than the choline and chlorine ions, naively a higher mobility and, thus, higher conductivity could be expected for these ions. However, again different glass temperatures for the choline-chloride and Li-salt-based systems seem to at least partly explain the much lower $\sigma_{dc}$ of the latter, especially at low temperatures. The small size of the lithium ion causes a higher ionic potential (ratio of charge to ion radius) than for both the cation and anion in choline-chloride. This should give rise to stronger inter-ionic interactions and also enhance the interactions between the Li$^+$ ions and the HBD molecules. It seems that this causes a strengthening of the local, transient structural network within the liquid, which leads to higher viscosity at a given temperature and, thus, to an increased glass-transition temperature. Of course, the stronger interactions of the lithium ions amongst each other and with the other constituents also should directly reduce their mobility, without invoking a viscosity variation, which may explain the observed persistence of the conductivity differences up to high temperatures, far above $T_g$. A reduction of ionic conductivity due to the replacement of larger by smaller ions is a well-known phenomenon and was, e.g., reported for ionic liquids in Refs. 51 and 67. One has to clearly state here, that dielectric spectroscopy, of course, principally cannot distinguish between conductivity contributions from the cations and the anions and the latter certainly also contribute to the observed $\sigma_{dc}$. However, the results of the present work are consistent with the assumption that the Li cations play a significant role in the charge transport in these systems. This is in accord with the findings of rather high relative mobility of the lithium ions deduced for the urea system from molecular dynamics simulations,[13,68] and makes these lithium DESs promising candidates for lithium-based batteries.[15,16]

Just as previously found for ethaline, glyceline, and reline,[23] $\sigma_{dc}(T)$ of the present lithium-salt DESs can be reasonably well fitted by a modification of the empirical Vogel-Fulcher-Tammann (VFT) law[69,70,71] [lines in Fig. 6(a)]:

$$\sigma_{dc} = \sigma_0 \exp\left[\frac{-D_\sigma T_{VF\sigma}}{T - T_{VF\sigma}}\right] \qquad (1)$$

Here $\sigma_0$ is a prefactor, $D_\sigma$ is the so-called strength parameter, which quantifies the deviations from Arrhenius behavior,[72] and $T_{VF\sigma}$ is the Vogel-Fulcher temperature where $\sigma_{dc}$ approaches zero. The resulting fit parameters are listed in Table I.

TABLE I. Glass-transition temperatures as determined from DSC measurements and parameters obtained from the VFT fits of $\sigma_{dc}(T)$ and $\langle\tau\rangle(T)$ shown in Fig. 4 and the fragility parameter calculated from $D_\tau$.

|  | $T_g$ (K) | $T_{VF\sigma}$ (K) | $D_\sigma$ | $\sigma_0$ ($\Omega^{-1}$cm$^{-1}$) | $T_{VF\tau}$ (K) | $D_\tau$ | $\tau_0$ (s) | $m$ |
|---|---|---|---|---|---|---|---|---|
| LiOTf/EG | 175 | 122 | 12.4 | 5.3 | 128 | 11.4 | 4.8×10⁻¹³ | 68 |
| LiOTf/Gly | 208 | 154 | 11.5 | 22 | 157 | 12.3 | 5.2×10⁻¹⁵ | 64 |
| LiTFSI/urea | 239 | 208 | 4.1 | 1.4 | 203 | 5.2 | 7.9×10⁻¹⁴ | 129 |

Figure 6(b) shows an Arrhenius plot of the average reorientational $\alpha$-relaxation times $\langle\tau\rangle(T)$ (closed symbols), calculated from the parameters of the CD function[73] which was used for describing the $\alpha$ relaxation in the fits of the dielectric spectra. The open symbols again denote the results for the corresponding choline-chloride-based DESs.[23,25] Just as for $\sigma_{dc}(T)$, quantifying the translational ionic dynamics [Fig. 6(a)], the temperature dependence of $\langle\tau\rangle$, related to the dipolar reorientational dynamics, exhibits marked deviations from Arrhenius behavior, typical for glass-forming liquids.[17,18,37,74,75] The relaxation-time curves of Fig. 6(b) roughly appear like mirror images of the $\sigma_{dc}(T)$ curves in Fig. 6(a) and show corresponding variations, with LiTFSI/urea having the highest and LiOTf/EG the lowest $\langle\tau\rangle$ values among the lithium systems at a given temperature. The Li-salt systems exhibit systematically longer relaxation times than the choline-chloride ones with the same HBDs [cf. closed and open symbols in Fig. 6(b)]. Again, all these variations may partly be ascribed to the different glass temperatures of these systems, having in mind the simple picture of asymmetric particles rotating within a viscous medium. In general, a comparison of Figs. 6(a) and (b) reveals that both the ionic translational and the dipolar reorientational dynamics are at least roughly coupled. Similar coupling was previously stated for glyceline, ethaline, and reline[23] and very recently demonstrated to be mainly mediated by the viscosity with certain deviations for reline.[34] It will be treated in more detail below.

As shown by the solid lines in Fig. 6(b), $\langle\tau\rangle(T)$ of the three investigated DESs can be well fitted by the VFT law, Eq. (2),[69,70,71,72] again typical for supercooled liquids.

$$\langle\tau\rangle = \tau_0 \exp\left[\frac{D_\tau T_{VF\tau}}{T - T_{VF\tau}}\right] \quad (2)$$

The resulting parameters are listed in Table I. The strength parameters and Vogel-Fulcher temperatures from the fits of $\sigma_{dc}(T)$ and $\langle\tau\rangle(T)$ are of comparable order (Table I), again pointing to a certain coupling of the ionic and reorientational dynamics. From $D_\tau$ the fragility parameter $m$ can be calculated (Table I), which is an often-used measure of the deviations of $\langle\tau\rangle(T)$ from Arrhenius behavior.[76] We obtain values of 68, 64, and 129 for LiOTf/EG, LiOTf/Gly, and LiTFSI/urea, respectively. Within the strong-fragile classification of glass-forming liquids,[72] this characterizes the present DESs as intermediate to fragile. In general, the fragility of a glass-forming ionic conductor has an influence on its room-temperature conductivity as demonstrated for ionic liquids in Ref. 77. In the present case, e.g., the significantly lower $D_\sigma$ of the urea system corresponds to a more pronounced bending in its $\sigma_{dc}(T)$ curve in Fig. 6(a). This leads to an enhancement of its conductivity at room temperature, which becomes comparable to that of LiOTf/Gly despite the higher glass temperature of LiTFSI/urea. The reason for the higher fragility of the urea systems is unclear. One may speculate that rather complex ion-HBD aggregations as found for reline[78] could also exist in LiTFSI/urea, leading to a more complex energy landscape and, thus, higher fragility.[79,80]

To estimate the glass-transition temperature, often the criterion $\langle\tau\rangle(T_g) \approx 100$ s is used. Extrapolating the fit curves in Fig. 6(b), we arrive at 172, 208, and 233 K for LiOTf/EG, LiOTf/Gly, and LiTFSI/urea, respectively. As mentioned above, from DSC experiments we have obtained 175, 208,[81] and 239 K, respectively (Table I). For the ethylene-glycole and glycerol systems, the agreement is reasonable. The 6 K lower $T_g$ from the $\langle\tau\rangle(T)$ data, found for LiTFSI/urea, could indicate slightly decoupled, enhanced reorientational dynamics at low temperatures. Further investigations are necessary to clarify this issue.

For the HBD glycerol, relaxation-time data are available in a broad temperature range as shown by the plusses in Fig. 6(b).[82] Remarkably, while the admixture of choline chloride to glycerol leads to an acceleration of reorientational motions especially at low temperatures, signified by a reduction of the relaxation times (open circles),[23] for LiOTf we find a slowing down (closed circles). The latter was also reported for the admixture of LiCl to glycerol.[83] In general, it is plausible that the addition of ions partially breaks up the intermolecular hydrogen-bond network of pure glycerol. Our results suggest that, for the relatively large ions in glyceline, this weakens the structural network, which leads to faster molecular dynamics, including the reorientational motions characterized by $\langle\tau\rangle$. In contrast, the much smaller radius of the lithium ions, causing a high ionic potential, seems to give rise to stronger interactions and a slowing down of the dynamics, roughly speaking by partly replacing the hydrogen bonds with stronger ionic bonds. While these simple considerations seem plausible, one should be aware that complex aggregations between the ions and HBD molecules as reported for the choline-chloride-based DESs[78,84,85,86] are likely to also exist in the Li-salt systems. This should affect both, the dipolar reorientational and the ionic translational [Fig. 6(a)] dynamics. Further work, e.g., using nuclear magnetic resonance measurements, are necessary to clarify the role of such aggregates for the molecular and ionic dynamics in this class of DESs.

For the HBD ethylene glycol, temperature-dependent relaxation-time data are available, too.[23,87,88] Those from Ref. 23 are shown by the crosses (×) in Fig. 6(b). Just as for glycerol, the addition of a lithium salt leads to considerable slowing down of the reorientational motions [cf. crosses and closed squares in Fig. 6(b)]. However, as already noted in Ref. 23, in this case, the admixing of choline chloride to ethylene glycol does not significantly alter the relaxation time (cf. crosses and open squares), in marked contrast to glycerol. In light of the discussion in the previous paragraph, this would imply that the bonding within the structural network of the HBD molecules has similar strength both in pure ethylene glycol and ethaline. Notably, the ethylene glycol molecule only comprises two OH groups instead of three in glycerol. In the pure HBD, this should lead to a qualitatively different network that is less three-dimensionally interconnected than for glycerol. This weaker network seems to be less affected by the addition of large ions. Finally, for pure urea, only a single relaxation time in the melt at 406 K was reported in literature[29] precluding any statements about the influence of ion admixture.

## C. Coupling of reorientational and translational dynamics

In Ref. 23, for ethaline and glyceline nearly perfect coupling of the dipolar reorientational dynamics and the translational ion dynamics was evidenced by the identical temperature dependences of $\langle\tau\rangle$ and $\rho_{dc} = 1/\sigma_{dc}$, implying $\langle\tau\rangle \propto \rho_{dc}$.[89] In contrast, in reline weak but significant decoupling showed up. For the present Li-salt-based DESs, this coupling is examined in Fig. 7, where, for each material, the logarithms of both quantities are shown within the same frame, thereby ensuring the same number of decades on the two ordinates. Then, by properly choosing the starting values of the y-axes, a perfect match of the two curves would indicate direct proportionality of both quantities. Figure 7 reveals that $\langle\tau\rangle$ and $\rho_{dc}$ of the present DESs in general exhibit similar temperature dependence as already suggested by a comparison of Figs. 6(a) and (b). However, in contrast to ethaline and glyceline,[23,34] especially for LiOTf/Gly and LiOTf/EG small but significant decoupling effects show up while for the urea-based sample a better match of both quantities can be achieved in this way. In Fig. 7, we adjusted the y-axes to reach an agreement at the highest covered temperature, because at high temperatures decoupling effects usually are least pronounced. $\langle\tau\rangle(T)$ and $\rho_{dc}(T)$ in LiOTf/Gly and LiOTf/EG show growing differences with decreasing temperature, finally reaching about one decade. In this respect, these materials resemble the behavior previously found for reline.[23] Interestingly, for the latter very recent viscosity measurements in a broad temperature range revealed that the reorientational motions are well coupled to the viscosity while the ionic motions are decoupled from both dynamics.[34] Corresponding viscosity measurements are necessary to clarify whether a similar scenario applies for the present Li-salt DESs and to help clarifying the better coupling found for LiTFSI/urea.

In Ref. 34, for glyceline, ethaline, and reline nearly perfect reorientation-viscosity coupling was found. If it also applies to the present systems, the crosses in Fig. 7 characterizing the reorientational dynamics also provide an estimate of the temperature dependence of the viscosity. Then the results of Fig. 7 imply that at low temperatures the ionic mobility of LiOTf/EG and LiOTf/Gly is enhanced compared to the naive picture of a sphere that translationally moves within a viscous medium. In any case, the decoupling effects observed in Figs. 7(a) and (b) demonstrate that a revolving-door-like scenario,[20,21,90] where the rotation of the dipoles opens up paths for the translational motions of the ions, is not the dominant charge-transport process in these two materials. This mechanism should lead to good coupling of reorientational and ionic translational motions (i.e., $\langle\tau\rangle \propto \rho_{dc}$) as found for plastic crystals,[20,21] various ionic liquids,[22] and also for glyceline and ethaline[23,34] but which is not fulfilled for two of the present DESs (Fig. 7). It should be noted that for many plastic crystals the applicability of the revolving-door mechanism is quite evident,[20,21,90] and for ionic liquids it also seems likely.[22] However, for the DESs glyceline and ethaline the results can also be explained by rotation-translation coupling via the viscosity.[34] In general, deviations from $\langle\tau\rangle \propto \rho_{dc}$ exclude dominant revolving-door charge transport but the validity of this proportionality in a material may also be explained without invoking this mechanism.

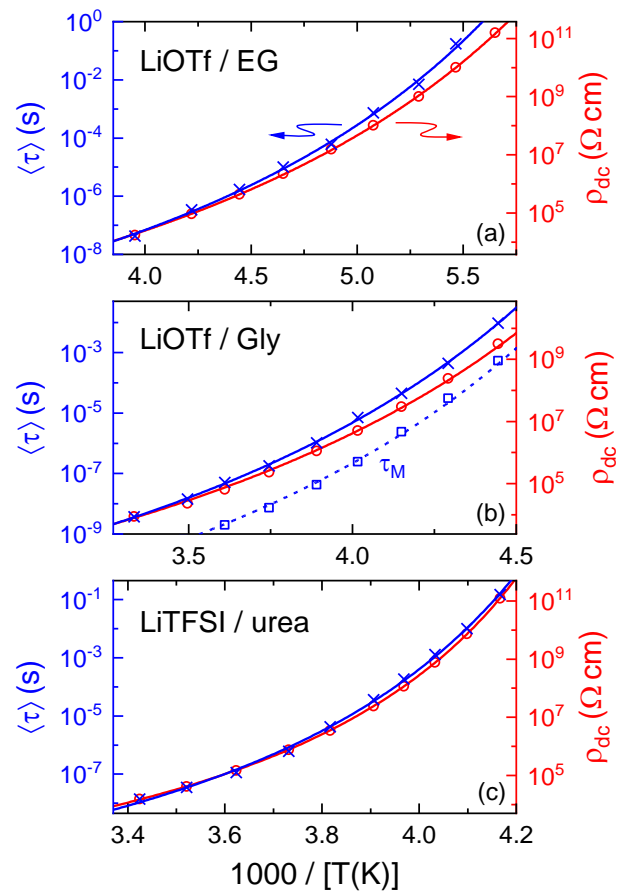

FIG. 7. Arrhenius plots of the average $\alpha$-relaxation times (crosses, left ordinates) and the dc resistivities (circles, right ordinates) of the investigated DESs. For each frame, the number of decades covered by the left and right ordinates is identical. Their starting values were adjusted to achieve a match of the two quantities at the highest investigated temperature. The soliid lines are fits with the VFT law (same as in Fig. 6), taking into account $\rho_{dc} = 1/\sigma_{dc}$. For LiOTf/Gly (b), in addition the relaxation times $\tau_M$, determined from the peak frequencies in the dielectric modulus spectra (Fig. 5) are shown (squares). The dashed line corresponds to the fit curve of the $\alpha$-relaxation time (upper solid line), shifted down by a factor of 22.

As mentioned above, for the choline-chloride-based DESs, good coupling of ionic and rotational motions was found for the HBDs ethylene glycol and glycerol, but some decoupling showed up for urea.[23] It is puzzling that in the present Li-salt-based DESs the situation is just the other way round, i.e., rather good coupling arises for the urea system and worse coupling for EG and glycerol (Fig. 7). The urea molecule differs by having hydrogen-bonding $NH_2$ and carbonyl groups instead of the hydroxyl groups of the alcohol HBDs, but it is unclear how this affects the coupling. It should also be noted that the present urea system has a different anion (TFSI) than the other two systems (OTf) which also may be the reason for the better coupling in this DES. The TFSI anion clearly is significantly bulkier than the OTf one which should affect cation and anion transport in some respect, e.g., requiring the opening of large gaps by the revolving HBD molecules for anion transport. While dielectric spectroscopy reveals intriguing differences of the urea systems, it cannot finally solve these questions and, e.g., nuclear magnetic resonance measurements should be performed to learn more about the underlying microscopic mechanism of the deviating behavior of the urea systems, both in the lithium- and choline-chloride-based DESs.

In Fig. 7(b), in addition the relaxation times $\tau_M$, determined from the peak frequencies in the dielectric modulus spectra (Fig. 5), are included (squares).[91] In an inconsiderate application of the modulus formalism, $\tau_M$ would correspond to the conductivity relaxation time, characterizing ionic mobility. However, as discussed in detail above, in the present case of an ionic conductor with strong reorientational degrees of freedom, the main peak in the $M''$ spectra arises from the dipolar motions of the HBD. This is nicely corroborated by the fact that $\tau_M(T)$ in Fig. 7(b) reveals the same temperature dependence as the $\alpha$-relaxation time, i.e., $\tau_M(T) \propto \langle\tau\rangle(T)$ [cf. dashed line in Fig. 7(b)].[64] In contrast, $\tau_M(T)$ is clearly decoupled from $\rho_{dc}(T)$. The resistivity is inversely proportional to the charge-carrier density and mobility. If $\tau_M$ would indeed reflect the ionic mobility, the observed decoupling of $\tau_M(T)$ and $\rho_{dc}(T)$ would imply a strong temperature-dependent variation of the number density of the ions [by almost one decade, cf. Fig. 7(b)] which is unlikely. The other two investigated DES systems exhibit qualitatively similar behavior.

Finally, Fig. 8 shows the variation of the dc resistivity with the $\alpha$-relaxation time (obtained at different temperatures) for the investigated DESs. As noted in Ref. 23, for glyceline and ethaline $\rho_{dc} \propto \langle\tau\rangle$ is valid (upper dashed line with slope 1 in Fig. 8), in accord with the so-called Debye-Stokes-Einstein (DSE) relation.[92,93,94,95] In contrast, for reline a fractional power law was found, $\rho_{dc} \propto \langle\tau\rangle^\xi$ with $\xi = 0.93$ (lower dashed line). It reminds of the fractional DSE behavior reported for various glass-forming systems.[92,93] As already suggested by Fig. 7, the Li-salt-based DESs investigated in the present work also significantly deviate from the DSE relation. Indeed, like reline they follow fractional power laws with $\xi = 0.89$, 0.88, and 0.96 for LiOTf/EG, LiOTf/Gly, and LiTFSI/urea, respectively (solid lines in Fig. 8). Once again, the urea-based DES displays a different behavior, compared to the other two systems. In agreement with the rather good match of the two curves in Fig. 7(c), the exponent $\xi$ is close to one for this DES. Figure 8 reveals that, especially at low temperatures, LiOTf/Gly and LiTFSI/urea have clearly enhanced dc conductivity (reduced dc resistivity) for a given relaxation time and probably also for a given viscosity when assuming the mentioned reorientation-viscosity coupling.[34] However, one should be aware that their conductivities for a given *temperature* are lower than for the choline-chloride DESs. Notably, without the decoupling evidenced by the fractional DSE relation, $\sigma_{dc}$ of these systems would be even lower.

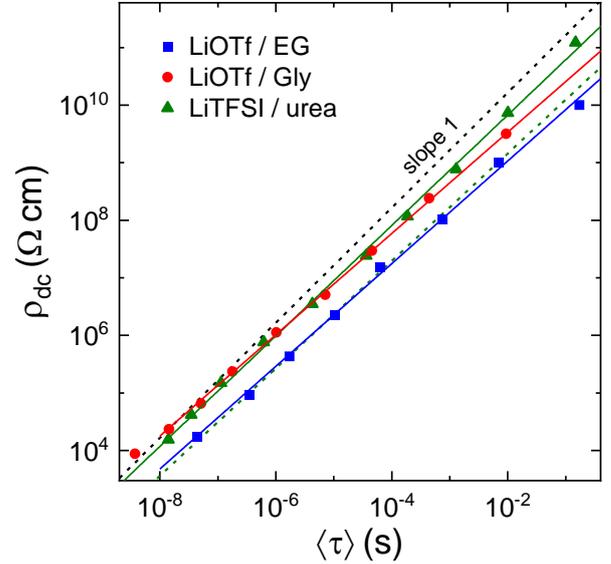

**FIG. 8.** Variation of the dc resistivities of the three investigated DESs with their reorientational relaxation times shown in a double-logarithmic plot. The upper dashed line with slope one indicates linear behavior, $\rho_{dc} \propto \langle\tau\rangle$, and represents the previously used fit curve describing the glyceline and ethaline data.[23] The solid lines are fractional power laws $\rho_{dc} \propto \langle\tau\rangle^\xi$, approximating the present systems. The lower dashed line shows the power law employed for fits of the reline data in Ref. 23.

## IV. SUMMARY AND CONCLUSIONS

In summary, we have performed broadband dielectric spectroscopy measurements on three DESs where the hydrogen-bond acceptor is a lithium salt. These measurements cover a broad dynamic range extending from the low-viscosity liquid around room temperature well into the supercooled-liquid state and finally approaching the glass state close to $T_g$. Just as for the previously investigated systems, composed of the same HBDs but containing choline chloride instead of a lithium salt, we find the signatures of a dipolar relaxation process revealing the characteristics of glassy freezing, in particular pronounced non-Arrhenius behavior. The HBDs represent the by far largest fraction of these DESs and the pure HBDs all have well-pronounced relaxation processes due to dipolar reorientations, whose relaxation times at high temperatures are of similar order as the present ones.[23,26,37,82,87,88] Therefore it is reasonable to assume that the detected relaxation process is predominantly due to reorientational motions of the HBD molecules. This is corroborated by the outcome of alternative analyses using the RBM or the modulus formalism, assuming purely translational relaxation processes. We find that, in general, for the investigated Li-salt-based DESs the reorientational dynamics is slower than for the corresponding choline-chloride systems and also slower than for the pure HBDs, especially at low temperatures. These findings and the detected strong relaxation-time differences between the three DESs can be rationalized by variations in the glass-transition temperature and/or the high ionic potential of the added lithium ions.

The temperature dependence of the technically relevant dc conductivities of these lithium DESs also reveals clear non-Arrhenius behavior, indicating that the translational ionic dynamics in principle is governed by the same glassy freezing as evidenced by the reorientational relaxation times. Analogous to the observed slowing down of the reorientational dynamics of the investigated lithium DESs with respect to the choline-chloride systems, their ionic translational dynamics, quantified by the conductivity, is also significantly reduced. This again can be in principle understood when considering the strong interactions of the lithium ions. Moreover, the detected conductivity differences between these three DESs also mirror the corresponding variations of their reorientational relaxation times. Their different $T_g$ and partly also fragility values qualitatively explain these results.

While there is a principle link of both the translational ionic and reorientational dipolar motions to the glassy dynamics, leading to similar temperature dependence, a direct comparison of the temperature-dependent relaxation times and dc resistivities reveals some significant deviations from a perfect coupling, in particular at low temperatures. This is especially obvious for LiOTf/EG and LiOTf/Gly, in marked contrast to ethaline and glyceline[23] containing choline-chlorine instead of the lithium salts. Therefore, a revolving-door-like mechanism seems not to be the dominant charge-transport mechanism in these materials. Just as previously reported for reline, we find a fractional DSE relation for the lithium DESs, which involves an enhancement of the ionic conductivity with decreasing temperature, finally reaching about one decade for LiOTf/EG and LiOTf/Gly. It thus seems likely that in these systems the small lithium ions find paths within the liquid structure enabling enhanced diffusion going beyond that expected for a viscous medium at low temperatures. It should be mentioned here that the observed decoupling could also reflect a temperature-dependent variation of the number of ion pairs, i.e. of ionicity, which would lead to a variation of conductivity via the number of free ions available for the charge transport. These effects seem to be less important for LiTFSI/urea where the mentioned decoupling is less pronounced. This could be related to the different hydrogen bonding in this urea DES or its larger LITFSI anions. We hope the present work will stimulate future nuclear magnetic resonance measurements aiming to clarify these open questions. Anyway, it should be noted that, in most of the investigated temperature ranges, the lithium DESs have conductivity values that are many decades lower than those of the corresponding choline-chloride systems with the same HBDs. As revealed by Fig. 6(a), fortunately (from an application viewpoint) these differences are strongly reduced close to room temperature which essentially can be traced back to variations in the VFT parameters describing the pronounced non-Arrhenius behavior of all these systems.

Overall, the results of the present work demonstrate the high relevance of the often-neglected glass-forming properties of DESs and of decoupling phenomena for the absolute values of their ionic dc conductivity. In addition to the present dielectric results, it certainly would be interesting to collect additional microscopic information on these materials, especially on the modification of the hydrogen-bond network by the added lithium ions and on the likely formation of ion-HBD clusters, using infrared spectroscopy and nuclear magnetic resonance measurements.

## ACKNOWLEDGMENTS

This work was supported by the Deutsche Forschungsgemeinschaft (grant No. LU 656/5-1). We thank Roland Böhmer for helpful discussions.

## DATA AVAILABILITY

The data that support the findings of this study are available from the corresponding author upon reasonable request.

___________________________________